\documentclass[aip,reprint]{revtex4-2}

\usepackage{graphicx}
\usepackage{float}
\usepackage{verbatim}
\usepackage{units}
\usepackage[usenames,dvipsnames]{color}
\usepackage{bm}
\usepackage{mathrsfs}
\usepackage{amsmath}
\begin{document}

\title{Electronic Structure and Epitaxy of CdTe Shells on InSb Nanowires} 


\author{Ghada Badawy}
\affiliation{Applied Physics Department, Eindhoven University of Technology, 5600 MB Eindhoven,~The Netherlands}
\author{Bomin Zhang}
\affiliation{University of Pitsburgh, Pittsburgh, PA 15260, USA}
\author{Tomá\v{s} Rauch}
\affiliation{Institut für Festkörpertheorie und -optik, Friedrich-Schiller-Universität Jena, 07743 Jena, Germany}
\author{Jamo Momand}
\affiliation{Zernike Institute for Advanced Materials, University of Groningen, 9747 AG Groningen, The Netherlands}
\author{Sebastian Koelling}
\affiliation{Department of Engineering Physics, Ecole Polytechnique de Montéal, C.P. 6079, Succ. Centre-Ville, Montréal, Québec, Canada H3C 3A7}
\author{Jason Jung}
\affiliation{Applied Physics Department, Eindhoven University of Technology, 5600 MB Eindhoven,~The Netherlands}
\author{Sasa Gazibegovic}
\affiliation{Applied Physics Department, Eindhoven University of Technology, 5600 MB Eindhoven,~The Netherlands}
\author{Oussama Moutanabbir}
\affiliation{Department of Engineering Physics, Ecole Polytechnique de Montéal, C.P. 6079, Succ. Centre-Ville, Montréal, Québec, Canada H3C 3A7}
\author{Bart J. Kooi}
\affiliation{Zernike Institute for Advanced Materials, University of Groningen, 9747 AG Groningen, The Netherlands}
\author{Silvana Botti}
\affiliation{Institut für Festkörpertheorie und -optik, Friedrich-Schiller-Universität Jena, 07743 Jena, Germany}
\author{Marcel A. Verheijen}
\affiliation{Eurofins Material Science Netherlands B.V., High Tech Campus 11, 5656 AE Eindhoven, The Netherlands}
\author{Sergey M. Frolov}
\affiliation{University of Pitsburgh, Pittsburgh, PA 15260, USA}
\author{Erik P.A.M. Bakkers}
\affiliation{Applied Physics Department, Eindhoven University of Technology, 5600 MB Eindhoven,~The Netherlands}

\begin{abstract}
Indium antimonide (InSb) nanowires are used as building blocks for quantum devices because of their unique properties, i.e., strong spin-orbit interaction and large Landé g-factor. Integrating InSb nanowires with other materials could potentially unfold novel devices with distinctive functionality. A prominent example is the combination of InSb nanowires with superconductors for the emerging topological particles research.
Here, we combine the II-VI cadmium telluride (CdTe) with the III-V InSb in the form of core-shell (InSb-CdTe) nanowires and explore potential applications based on the electronic structure of the InSb-CdTe interface and the epitaxy of CdTe on the InSb nanowires. We determine the electronic structure of the InSb-CdTe interface using density functional theory and extract a type-I band alignment with a small conduction band offset ($\, \leq 0.3$~eV). These results indicate the potential application of these shells for surface passivation or as tunnel barriers in combination with superconductors. In terms of the structural quality of these shells, we demonstrate that the lattice-matched CdTe can be grown epitaxially on the InSb nanowires without interfacial strain or defects. These epitaxial shells do not introduce disorder to the InSb nanowires as indicated by the comparable field-effect mobility we measure for both uncapped and CdTe-capped nanowires.
 
\end{abstract}

\maketitle 

\section{Introduction}

Semiconductor nanowires with strong spin-orbit coupling and large Landé g-factor have unfolded novel research pathways in quantum transport ranging from spin phenomena~\cite{scherubl2016electrical, rossella2014nanoscale,liang2012strong,bandyopadhyay2014coherent} to quantum computing circuits~\cite{petersson2012circuit, larsen2015semiconductor,nadj2010spin}, and most recently the pursuit of topological particles~\cite{mourik2012signatures,deng2016majorana,das2012zero}. Indium antimonide (InSb) is among these semiconductors, with the highest bulk electron mobility, largest Landé g-factor, and strongest spin-orbit coupling compared to other III-V materials~\cite{nilsson2009giant,van2015spin}. Integrating these InSb nanowires with other materials holds great potential for realizing novel devices with unconventional functionality, yet it has been hampered in large part due to the large lattice constant of InSb that tends to complicate the realization of defect-free and strain-free heterostructures.

The II-VI material, cadmium telluride (CdTe), is an interesting material candidate as it nearly matches the large lattice constant of InSb with a lattice mismatch below 0.05\% at room temperature. Moreover, their thermal expansion coefficients are comparable~\cite{luna2020strategies}. Besides their structural compatibility, CdTe has a large bandgap compared to InSb. Therefore, combining InSb with CdTe is compelling for several device applications such as, quantum-well lasers, high electron mobility transistors, and infrared detectors~\cite{mackey1986chemical,zhang2020highly}. However, in spite of the nearly perfect lattice-match, growth of CdTe-InSb heterostructures remains complicated, due to preferential interface reactions which lead predominantly to the formation of an indium-tellurium rich interface region~\cite{mackey1987insb}. The formation of such layer is undesirable, since different compositions of this indium telluride compound have different lattice constants and bandgaps~\cite{mackey1987insb}. The ease of twin formation in CdTe crystals~\cite{farrow1981molecular,vere1983origins}, further complicates the realization of defect-free interfaces.

Here, we combine InSb and CdTe in a core-shell (InSb-CdTe) configuration for potential applications based on the electronic structure of the InSb-CdTe interface, the structural quality, as well as, epitaxy of the CdTe shells on the InSb nanowires. The electronic structure of the InSb-CdTe is investigated with density functional theory (DFT) calculations, where both the bandgaps and the band-edge alignment at the InSb-CdTe interface are extracted. In particular, we show that the electronic structure at the InSb-CdTe interface is well-suited for passivating the InSb surface by virtue of a type-I band alignment and for serving as a tunnel barrier when placed at the interface between the InSb nanowire and a metal or superconductor owing to a small conduction band offset of roughly 0.3~eV. In the context of engineering topological superconductors using superconducting-semiconducting nanowire hybrids, such a CdTe tunnel barrier could potentially minimize disorder and address the strong-coupling issue between the superconductor and the nanowire. Disorder in these nanowire hybrids mimics the signatures of topological particles~\cite{ahn2021estimating} and an overly strong coupling is deemed to overwhelm the intrinsic properties of the semiconducting nanowire and possibly render topological superconductivity inaccessible~\cite{cole2015effects,cole2016proximity}. Using CdTe shells at the interface between the nanowire and a superconductor with the extracted conduction band offset could modulate the superconductor-semiconductor coupling strength. In addition to the electronic structure of the InSb-CdTe interface, the crystal quality of the interface and of the CdTe shells is crucial for device applications. In particular, defected shells could induce strain in the InSb nanowire and introduce new sources of disorder that could impair the performance of the core-shell nanowires~\cite{kavanagh2011transport,popovitz2011inas}. Therefore, we also demonstrate the growth of defect-free, epitaxial CdTe shells with a smooth and abrupt interface to InSb nanowires. The growth parameters have been optimized to suppress interface reactions and therefore the formation of interface layers is  suppressed entirely. Furthermore, the CdTe shell inhibits the InSb nanowire surface from oxidizing which eliminates the need for exposing the InSb nanowires to harsh chemicals to remove surface oxides for device fabrication. As discussed in Section II-B, we determine that the CdTe shells are chemically stable against oxidation, where after a period of at least three weeks they remain oxide-free. This property facilitates device fabrication, specifically in devices where the CdTe needs to be contacted, such as tunnel barrier devices. In this case, the metal or superconductor can be directly deposited on the CdTe without having to expose the CdTe to etchants and harsh chemicals.
The quality of the grown shells is corroborated by transport measurements, where we obtain comparable electron mobility values for both bare, uncapped and CdTe-capped InSb nanowire, thus confirming that these shells do not introduce additional disorder to the nanowires.

\begin{figure*}[t!]
	\centering
	\includegraphics[scale=1]{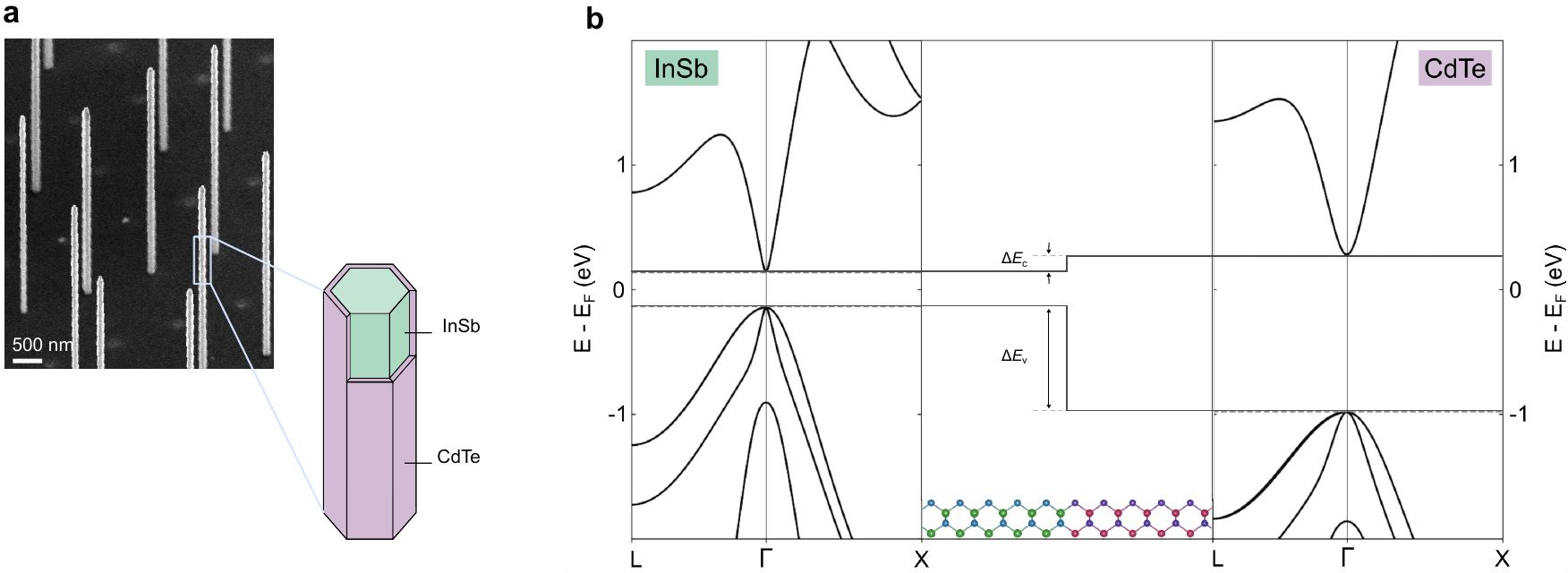}
	\caption{\textbf{CdTe-capped InSb nanowires and the electronic structure of the InSb-CdTe system.}\textbf{ a.} CdTe shells on InSb nanowires imaged at a 30$^{\circ}$-tilt with a scanning electron microscope (SEM). The schematic highlights the InSb core (green) surrounded by a CdTe shell (purple). For illustration purposes, the CdTe shell is removed on two facets to expose the InSb.\textbf{ b.} On the left and right are the bulk bandstructures of InSb and CdTe, respectively, as obtained from density functional theory. In the central panel, the supercell composing the InSb-CdTe interface is shown with the band alignment. The potential difference between both materials is extracted from the averaged local electrostatic potential and is used for the alignment of both bandgaps at the interface. The band edge alignment has a conduction band offset $\Delta E_\text{c}$~=~0.15~eV and a valence band offset $\Delta E_\text{v}$~=~$-0.84$~eV. }
	\label{dft_fig}
	\centering
\end{figure*}

\section{Results}

\subsection{Electronic structure of the InSb-CdTe interface}

The electronic structure across the interface between the InSb and the CdTe is characterized using \textit{ab initio} DFT calculations. The atomic structure of the interface (i.e., the supercell) along with selected results from the DFT calculations are graphically presented in Figure~\ref{dft_fig}b. The supercell, displayed in the middle panel of Figure~\ref{dft_fig}b, is created based on experimental inputs extracted from the structural and composition analysis of the grown shells. Accordingly, the interface is oriented perpendicular to the $<110>$ crystal direction, in line with the CdTe coverage of the six equivalent \{220\} facets, which outline the hexagonal cross-section of an InSb nanowire (Figure~\ref{dft_fig}a). The supercell also accounts for the polarity of the InSb-CdTe interface, in accordance with the analysis discussed in Section B, where Cd takes the position of In, and Te that of Sb. The nearly perfect lattice match allows for the assumption that both InSb and CdTe have the same lattice constant. This is substantiated by our observations of the absence of strain in the grown shells (see Section B). Therefore, an ideal interface is assumed with the lattice constant of the supercell set to the experimental value of CdTe. Additional calculations, details, and results are provided in Section~I of the supplementary information.

The results presented in Figure~\ref{dft_fig}b show the individual bulk bandgaps of both InSb and CdTe, as well as, the band alignment at the interface of the heterostructure, i.e., the supercell. The potential difference at the interface, calculated from the local electrostatic potential of the supercell, is used to shift the InSb and CdTe bands, yielding the shown alignment. This alignment is commonly referred to as type-I, with the InSb band-edges lying between those of CdTe. From the band-edge alignment a conduction band offset $\Delta E_\text{c}$ = 0.15~eV and valence band offset $\Delta E_\text{v}$~=~$-0.84$~eV are extracted. Since the bandgap of CdTe is slightly underestimated\textemdash compared to experimental values\textemdash $\Delta E_\text{c}$ is correspondingly underestimated by roughly 0.2~eV. In contrast, the valence band offset is in good agreement with reported experimental values and theoretical calculations~\cite{mackey1987insb,wang2018band,tersoff1986band,hinuma2014band}. For the conduction band offset, however, there has been little consensus on its accurate value. Estimations based on the electron affinity rule give an offset of roughly 0.3~eV~\cite{van1984properties,williams1985mbe}. 

\begin{figure*}[t!]
    \centering
    \includegraphics[scale=1]{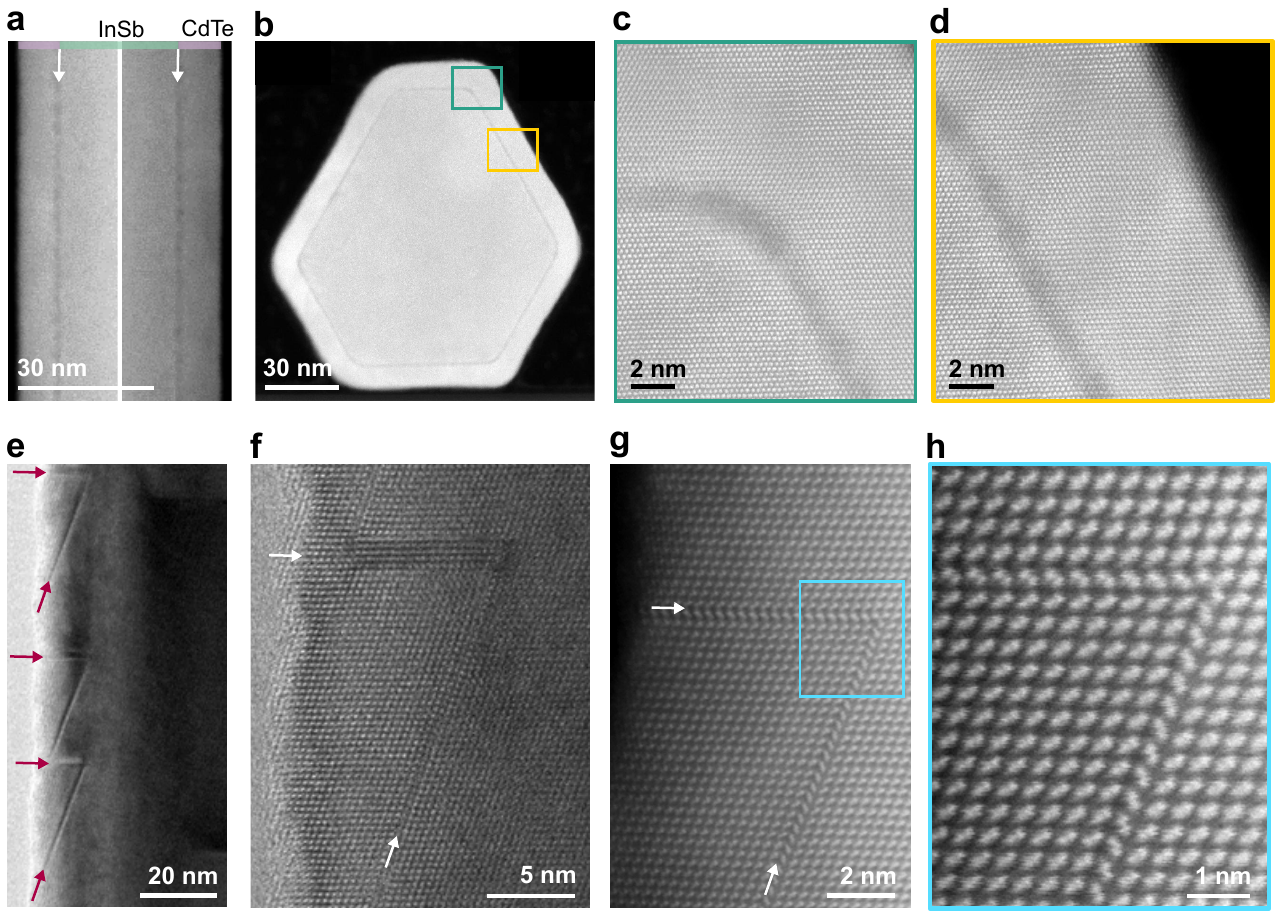}
    \caption{\textbf{The impact of interfacial oxides.}\textbf{ a.} STEM images of the two opposing sides of the same nanowire imaged along the $<112>$ zone axis show a dark layer between the CdTe shell and the InSb core, indicated by the arrows. The dark layer is attributed to InSb oxides, appearing darker because they are of a lower electron density.\textbf{ b.} A cross-section of a nanowire imaged along the $<111>$ zone axis shows this oxide layer is all-around, where high magnifications of the interface in\textbf{ c.} and\textbf{ d.} indicate that the limited thickness of this oxide layer still allows for epitaxy between InSb and CdTe.\textbf{ e.} Despite epitaxy, defects in the CdTe shell are detectable (indicated by arrows) with high-resolution TEM along the $<110>$ zone axis. Two directions of planar defects are present, both parallel to \{111\}~B planes: orthogonal to the long axis and at roughly 19$^{\circ}$ from the long axis. Different combinations of twinned layers in each direction occur.\textbf{ f.} In the orthogonal direction four twinned layers are visible and in the inclined direction only a single pair of twin boundaries.\textbf{ g.}~Pairs of twin boundaries in each direction. HAADF-STEM scans show that the defect starts at a specific point and then expands by two double \{111\} twin planes at a 71$^{\circ}$ angle.\textbf{ h.} A high-magnification scan of the starting point of the defect clearly reflects the interrupted crystal structure by the two twins. }
    \label{oxides_fig}
\end{figure*}


\subsection{Epitaxy of CdTe shells}\label{background}

The InSb nanowires studied here are grown using the vapor-liquid-solid technique on masked InSb (111)~B substrates from gold catalysts defined by electron-beam lithography, as detailed in~\cite{badawy2019high}, resulting in uniform arrays of nanowires (see Figure~\ref{dft_fig}a). In our first approach to prevent the InSb nanowires from oxidizing, we try transferring them from the metal-organic vapor phase epitaxy (MOVPE)\textemdash where they are grown\textemdash to the molecular beam epitaxy (MBE) cluster\textemdash where the CdTe is deposited\textemdash under nitrogen overpressure. However, this nitrogen environment with low levels of oxygen ($\, < 1$~parts per million) is not enough to inhibit the InSb nanowires from oxidizing. The presence of these oxides manifests as a dark contrast layer between the InSb core and the CdTe shell in scanning transmission electron microscopy (STEM) imaging (Figure~\ref{oxides_fig}a-d) and triggers the formation of defects (Figure~\ref{oxides_fig}e-h). The high defect density we observe in the CdTe shell of Figure~\ref{oxides_fig}e-h is consistent with reports on the presence of oxides on InSb surfaces affecting the quality of grown CdTe layers~\cite{wood1984microstructural}.
The nanowire cross-section of Figure~\ref{oxides_fig}b with a uniformly thick CdTe shell of 7~nm indicates that this (dark-contrast) oxide layer at the core-shell interface is present all-around the InSb nanowire. Yet, high-magnification images of the interface close to a corner and parallel to a facet (Figure~\ref{oxides_fig}c,d) signify that the oxide layer is not fully formed as evidenced by the continuation of atomic columns locally from the core to the shell. In addition, the oxide layer is not thick enough to disrupt an epitaxial connection. Despite the limited thickness of this oxide layer, it is enough to trigger twin defects in the CdTe shell. As shown in Figure~\ref{oxides_fig}e-h, the twin defects are oriented parallel to \{111\} planes, that is perpendicular or inclined by approximately 19$^{\circ}$ with respect to the long axis of the nanowire. In either direction, twinned layers come in pairs, one or multiples, such that the lattice orientation remains unchanged. Furthermore, high-angle annular dark field (HAADF) STEM scans (Figure~\ref{oxides_fig}g,h) show that the defect expands from a given point by double \{111\} twin planes that form a 71$^{\circ}$ angle.

Accordingly, to prepare the InSb nanowires for defect-free epitaxial CdTe shells it is essential to remove any surface oxides. In this work, the InSb nanowires are cleaned with atomic hydrogen in an MBE chamber prior to the growth of the CdTe shells. The reactive atomic hydrogen species are known to be effective in eliminating surface oxides from III-V semiconductors without changing stoichiometry or inducing roughness~\cite{sugata1988gaas, khatiri2004atomic}. Other than preserving the pristine quality of the InSb nanowires, atomic hydrogen cleaning in an MBE system has the advantage of providing ultra-high vacuum conditions throughout the entire process from cleaning to CdTe growth, thus ensuring that the nanowires remain oxide-free after cleaning.

The parameters used for oxide removal with atomic hydrogen need to be carefully tuned, as the choice of parameters can be detrimental to the quality of both the InSb nanowires and the deposited CdTe shells. On the one hand, relatively low temperatures ($\,< 200^{\circ}$~C) are not enough to completely remove the native oxides, and thus lead to the growth of defected CdTe shells.
On the other hand, higher temperatures ($\, > 300^{\circ}$~C) compromise the quality of the InSb nanowires by inducing surface roughness (see Figure~S1 in the supplementary information). Optimized cleaning parameters enable the growth of defect-free epitaxial shells, free of interfacial oxides, as presented in Figure~\ref{uniformCdTe_fig}.
Details on atomic hydrogen cleaning are outlined in Section II of the supplementary information.

\begin{figure}[t!]
	\centering
	\includegraphics[scale=1]{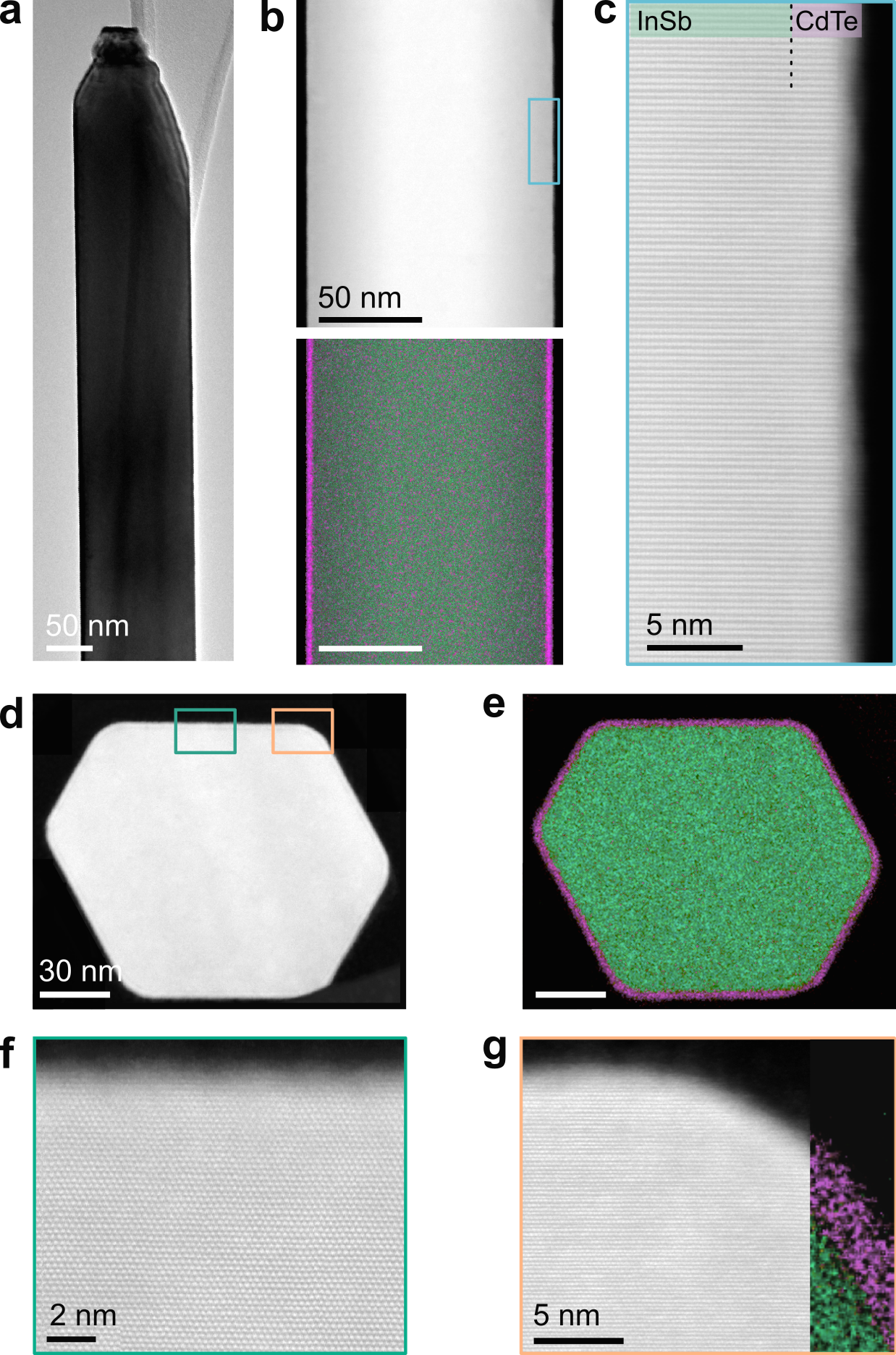}
	\caption{\textbf{Growth of epitaxial uniform CdTe shells.}\textbf{ a.} A representative CdTe-capped InSb nanowire imaged with transmission electron microscopy along the $<112>$ zone axis.\textbf{ b.} A HAADF scan of a nanowire segment ($<112>$ zone axis) along with an EDX map of a uniform CdTe shell (purple) around an InSb nanowire core (green).\textbf{ c.} A zoom-in on the framed region conveys epitaxy from core to shell, where the interface is virtually indistinguishable.\textbf{ d.}~A cross-sectional HAADF scan accompanied by\textbf{ e.} an EDX map shows a 2.5~nm full shell and two high magnification images, \textbf{ f.}~parallel to a facet and\textbf{ g.} along the corner. Both reveal the abrupt interface and epitaxy between the InSb and the CdTe. In\textbf{ g.}, an EDX map is overlaid to identify the CdTe and the InSb.}
	\label{uniformCdTe_fig}
	\centering
\end{figure}

Following the oxide removal, the CdTe is deposited from two separate cells, a Cd effusion cell and a Te cracker cell. During growth, the base pressure in the chamber is around $1 \times 10^{-9}$ Torr. Three key considerations are necessary to suppress the formation of interface reactions which lead to Te-rich interface layers. First, the nanowires are exposed to a Cd flux for three minutes prior to the introduction of Te. Second, growth proceeds under Cd-rich conditions with a II/VI flux ratio~=~3. We note that only a high II/VI flux ratio is not sufficient to hinder the formation of Te-rich compounds at the core-shell interface. Particularly, shells deposited without pre-exposure to Cd exhibit a Te-rich interface layer, as measured with atom probe tomography (see Section~III of the supplementary material). These Cd-rich conditions do not affect the shell stoichiometry because of the low sticking coefficient of Cd and its high vapor pressure\textemdash four orders of magnitude higher than Te~\cite{mackey1987insb}. Third, relatively low growth temperatures are used ($T \approx 120 - 150^{\circ}$~C) to ensure interdiffusion processes do not occur between the InSb and the deposited CdTe. Moreover, to promote the growth of smooth CdTe shells, low Cd and Te fluxes are used, yielding a growth rate of $\approx 0.002$~monolayer/second. Higher growth rates result in defected and rough shells. While higher temperatures enhance selectivity owing to a longer diffusion length of the adatoms\textemdash~evident by the absence of CdTe deposition on the silicon nitride mask covering the nanowire substrate\textemdash they lead to thermal etch pits in both the InSb nanowires and substrate. The release of elemental antimony from InSb surfaces at elevated temperatures has been reported and can be minimized with an antimony overpressure~\cite{park2014effects}. The thermal etch pits formed in the nanowires compromise the structural integrity of the nanowires causing them to bend (Figure~S3). 

\begin{figure*}[t!]
    \centering
    \includegraphics[scale=1]{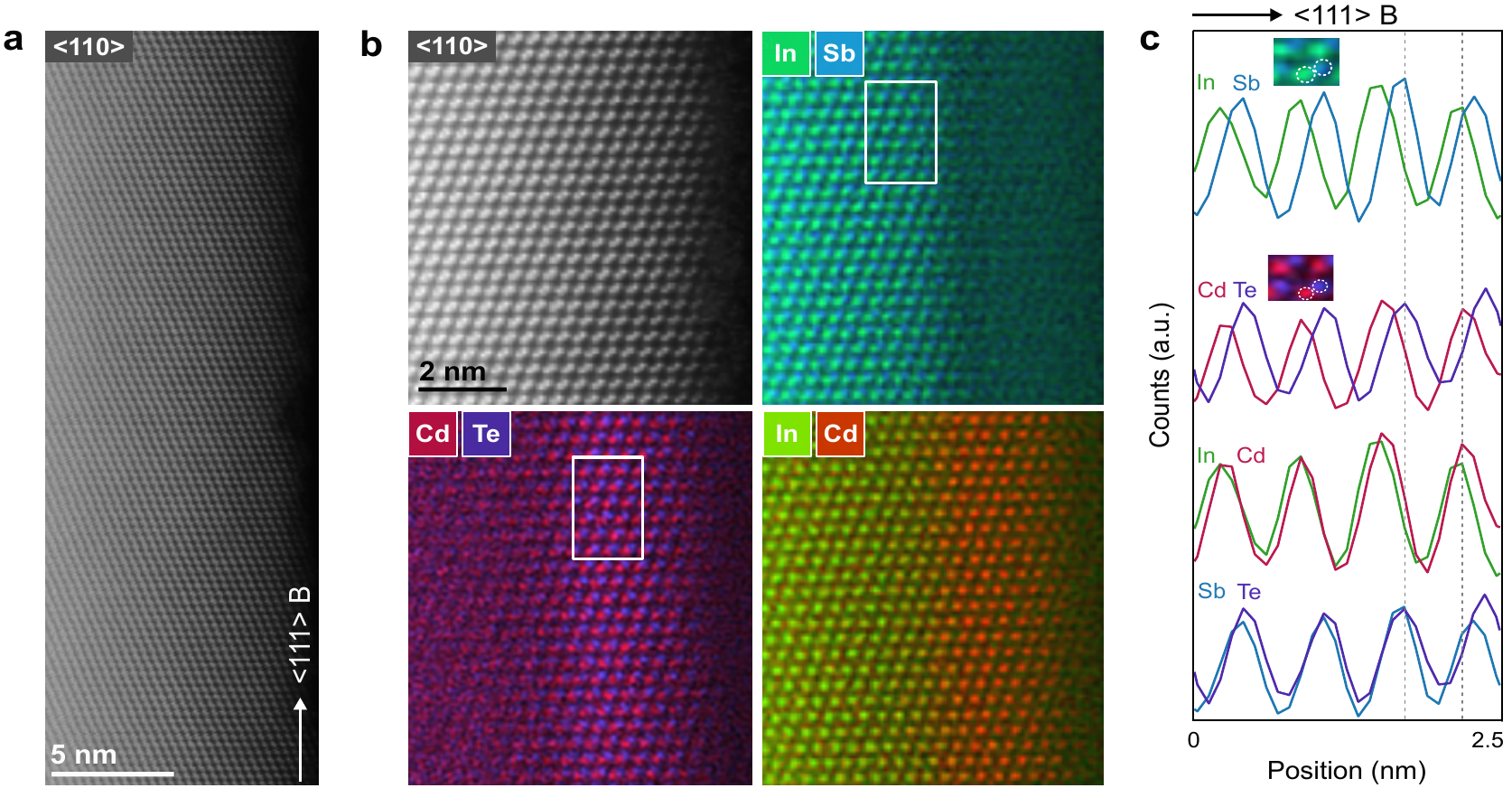}
    \caption{\textbf{Atomic structure and composition of the CdTe shells.}\textbf{ a.} An InSb nanowire covered with a 2.5~nm CdTe shell imaged with HAADF-STEM along the $<110>$ zone axis, as specified.\textbf{ b.} A high magnification scan of\textbf{ a.} demonstrates the ABC stacking of the zinc blende crystal of both InSb and the epitaxial CdTe. Accompanying atomic resolution EDX maps reveal the positions of the elements composing the CdTe shell and InSb core along the $<111>$~B growth direction.\textbf{ c.} Framed regions in\textbf{ b.} indicate along which areas the atomic profiles are taken. Integrating over the region yields the projection of the atomic positions on one line along the $<111>$~B. Inset are two high magnification images of the atomic resolution EDX maps, InSb and CdTe, to highlight the atomic positions of the elements in the core and shell. The atomic resolution EDX maps and the atomic profiles display the polarity of the shell with respect to the core, with Cd taking the position of In and correspondingly, Te that of Sb. This is further demonstrated by the absence of a shift in the line profiles of In (Sb) and Cd (Te), as highlighted by the vertical dashed lines.}
    \label{atomicEDX_fig}
\end{figure*}

Accordingly, low growth temperatures and fluxes are used to drive the growth of defect-free epitaxial CdTe shells. For instance, Figure~\ref{uniformCdTe_fig}a shows an overview bright field transmission electron microscopy (BFTEM) image of such an InSb-CdTe core-shell nanowire. The virtually indiscernible interface between the InSb and the CdTe in HAADF-STEM imaging, in Figure~\ref{uniformCdTe_fig}b,c, relates not only to the absence of any interface layers but also to the similar atomic numbers of all the elements\textemdash indium, antimony, cadmium, and tellurium\textemdash composing both the core and shell. Accordingly, as presented in Figure~\ref{uniformCdTe_fig}b, energy dispersive x-ray (EDX) spectroscopy mapping is used to deduce the thickness of the grown shells. The thickness of this particular shell is 2.5~nm and is uniform along the length of the entire nanowire. Moreover, Figure~\ref{uniformCdTe_fig}c conveys the defect-free epitaxy persisting from core to shell. Cross-sectional studies of the nanowire allow for the investigation of the shell quality orthogonal to the long axis, as presented in Figure~\ref{uniformCdTe_fig}d-g. The accompanying EDX map illustrates the uniform CdTe shell thickness on all six facets. This uniform, full shell is enabled by rotating the substrate during the CdTe growth.
High resolution imaging along the $<111>$ zone axis confirms the high-quality and defect-free epitaxy in the middle of a facet and at a corner. We emphasize that there is no visible contrast between the core and the shell demonstrating that there is abrupt epitaxy between InSb and CdTe without an interfacial layer.

This epitaxy is also clearly detectable in the high-magnification scans taken along the $<110>$ zone axis of the HAADF scanning mode presented in Figure~\ref{atomicEDX_fig}a, where the zinc blende structure of both the InSb and the CdTe is recognizable. An atomic-resolution composition mapping of the shell obtained with atomic-resolution EDX in Figure~\ref{atomicEDX_fig}b conveys the positions of the elements composing the core and the shell. These mappings reveal that the CdTe copies the polarity of the InSb nanowire. In particular, the (111)~B layers orthogonal to the growth direction are terminated by antimony atoms, in line with the (111)~B substrate orientation. Correspondingly, the (111)~B planes of the shell are terminated by tellurium atoms and cadmium takes the position of indium (Figure~\ref{atomicEDX_fig}b). This polarity is further demonstrated by atomic profiles taken along the InSb core and the CdTe shell where the projection of the atomic positions on one line along the $<111>$~B direction denotes the spatial overlap between In and Cd, and similarly Sb with Te in Figure~\ref{atomicEDX_fig}c. 

Close examination of the shells reveals that they do not oxidize (scans are taken at least three weeks after shell growth) evident by persistent atomic columns that are not bound by a low-electron density material or amorphous structure. The inertness of CdTe stands in stark contrast to InSb, which readily oxidizes at very low levels of oxygen. The susceptibility of InSb to oxidation highlights the importance of these shells since they effectively hinder the formation of surface oxides around the InSb nanowires.

The structural quality of these shells is assessed for strain. As expected, because of the near lattice-match, the shells are relaxed as confirmed by strain mapping provided in Figure~S4 of the supplementary information. A relaxed shell implies a large critical thickness, on the one hand, consistent with defect-free shells for the range of studied thicknesses up to 15~nm. On the other hand, the non-strained interface further confirms the absence of interface layers, since the commonly formed interfacial indium- and tellurium-rich compounds have a lattice mismatch of roughly 5\% with InSb~\cite{mackey1987insb}.

Tuning the CdTe shell thickness is simply achieved by varying the growth time, where a linear approximation of the growth rate yields roughly 5.4~nm/hour. We investigate shells with thicknesses from 2.5~nm up to 12~nm. We note that for shell thicknesses greater than 5~nm very slight roughness is observed with TEM along the $<110>$ zone axis (Figure~S5). Imaging the same shell along the $<112>$ zone axis does not reveal this roughness because images are taken parallel to a roughly 100-nm-long nanowire facet. Thus, aggregated nanoscale roughness is projected in the image plane. In contrast, $<110>$ zone axis viewing yields an image of the corner between two facets, thereby exposing any atomic scale roughness at the corners. This roughness shows up in projection at edges orthogonal to a $<111>$ direction. Although the exact topography and features of this roughness cannot be extracted, it could be related to the tendency of CdTe to form \{111\} facets with increased layer thickness, as described in~\cite{neretina2006role,ribeiro2007low}.

Eventually, growth is terminated by closing both the Cd and Te shutters, thus cool down proceeds without any fluxes. As a matter of fact, cooling down under a Te flux at low growth temperatures, e.g., 120-150$^{\circ}$~C, results in the deposition of Te-rich CdTe globules on the nanowire facets as depicted in Figure~S6 of the supplementary information.


\subsection{Electric characterization of the InSb-CdTe core-shell nanowires}

\begin{figure}[t!]
	\centering
	\includegraphics[scale=1]{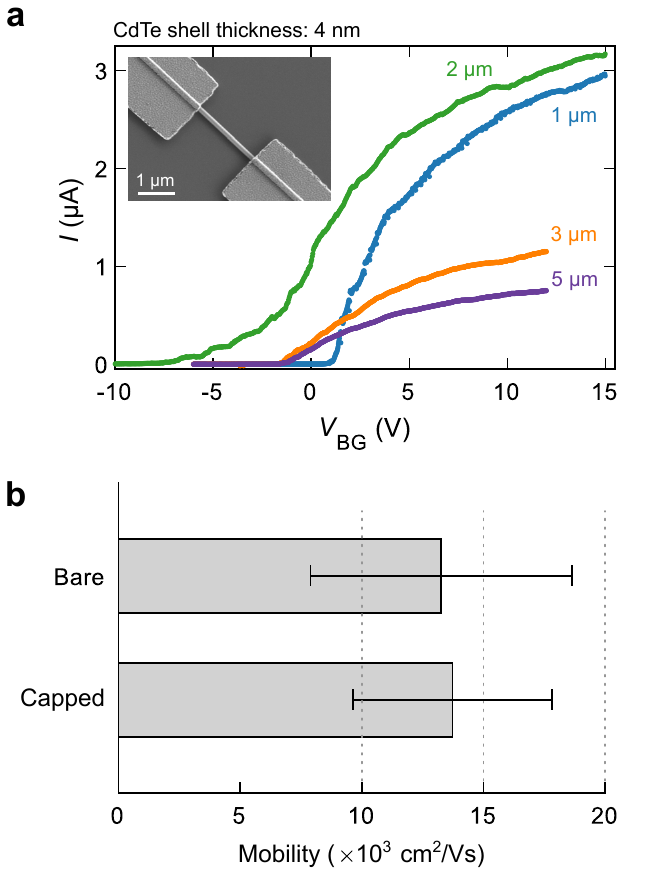}
	\caption{\textbf{Mobility in CdTe-capped nanowires.}\textbf{ a.} FET measurements at a bias voltage $V_{\text{dc}}$~=~10~mV for four core-shell nanowires (CdTe thickness = 4~nm) with channel lengths $L$~=~1, 2, 3, and 5~$\mu$m. Fitting these pinch-off curves yields mobility $\mu$ = 21.5, 26, 14, and $17 \times 10^3$~cm$^2$/Vs for 1, 2, 3, and 5~$\mu$m, respectively. Inset: SEM of a nanowire device.\textbf{ b.}~The average mobility for 4-nm CdTe capped and bare InSb nanowires. Mobility is averaged for 1- and 2-$\mu$m channel devices.}
    \label{avg_mob_fig}
	\centering
\end{figure}

The role of the epitaxial CdTe capping is evaluated by performing electron transport measurements on the core-shell nanowires. We note that we have not evaluated the tunnel barrier properties of the shell but studied the basic field-effect transistor (FET) characteristics of the core-shell wires. While there are limitations to mobility extraction from FET measurements in nanowires which could result in inexact mobility values~\cite{mcculloch2016avoid,choi2018critical}, FET measurements remain a commonly used technique for characterizing nanowires~\cite{ford2009diameter,gunawan2008measurement,dayeh2007high,gul2015towards}.
We perform FET measurements at 4~K to extract the mobility~$\mu$. A typical device is presented in the inset of Figure~\ref{avg_mob_fig}a. About 60 FET devices are fabricated with core-shell nanowire diameters of roughly 120~nm.  Degenerate p-doped silicon substrates covered with silicon oxide (SiO$_2$) and hafnium oxide (HfO$_x$) serve as a global back gate and titanium/gold contacts serve as the source-drain electrodes. Directly depositing the source-drain contacts onto the CdTe shells, for the studied shell thicknesses of 4 to 12~nm, results in an open-circuit, reflecting that these shells serve as good insulators. Therefore, prior to depositing the source-drain electrodes, the CdTe is etched locally by argon milling to make contact to the conducting InSb core. Additional details on device fabrication are provided in Section~VIII of the supplementary information. The source-drain contact separations are $L = 1, 2, 3,$ and 5~$\mu$m to ensure the long channel diffusive transport regime. Furthermore, these large separations ensure that any damage induced by the argon milling close to the source-drain contacts is eliminated and the measurements reflect the behavior of the segment between the contacts. Back-gate voltage sweeps $I (V_{\text{BG}})$ for the studied channel lengths and a fixed CdTe shell thickness of 4~nm are depicted in Figure~\ref{avg_mob_fig}a. Field-effect mobility values are extracted from fits of the pinch-off curves (Section~IX, supplementary information). The majority of the obtained mobility values are in the range of $1.0 - 2.7 \times 10^4$~cm$^2$/Vs and no significant difference is found with uncapped, bare InSb nanowires (Figure~\ref{avg_mob_fig}b). 
 
The equivalent mobilities for the CdTe shell and uncapped InSb nanowires indicate that the shell does not induce additional disorder to the InSb nanowires, thereby preserving the InSb nanowire properties. Possibly, thicker shells ($\, > 12$~nm) are required to passivate the nanowire surface and further confine the electrons in the nanowire, similar to InSb quantum wells, where relatively thick barrier layers ($\, > 50$~nm) are needed to achieve high mobility~\cite{lei2019quantum}. Nevertheless, thicker shells would not be compatible with a tunnel barrier at the interface between a semiconducting nanowire and a superconductor, seeing as the tunneling probability decreases exponentially with barrier thickness.

While the mobilities are comparable for both CdTe-capped and uncapped InSb nanowires, we observe a larger hysteresis between the forward and backward gate-voltage sweeps for the CdTe-capped wires compared to the uncapped wires (Figure S7 of the supplementary information). The origin of this hysteresis is unknown, nevertheless it could be attributed to point-defects at the interface and in the CdTe shells which are known to trap charges~\cite{choi2013native,wang2010hysteresis}. Point-defects do not possess a lattice structure in any dimension and accordingly are not visible in our TEM studies of the CdTe shells. The relatively low substrate temperatures used during the CdTe growth could possibly lead to the formation of point defects. They could also be formed during the atomic hydrogen cleaning of the InSb surface, resulting in point-defects at the InSb-CdTe interface.
We note that pumping the device space for 96~hours compared to 24~hours can slightly reduce the hysteresis and lead to a slight increase in the fitted mobility (supplementary material Figure S7). However, this slight improvement indicates that the hysteresis is likely dominated by attributes inherent to the interface and the CdTe shells, for instance, adsorbates related to point-defects and point-defects.


\section{Outlook}

We studied the InSb-CdTe material system in a core-shell nanowire configuration for potential applications in surface passivation and tunnel-barrier devices. We determined the suitability of these axial heterostructures for the proposed applications based on the electronic structure of the InSb-CdTe interface and the quality of the CdTe epitaxy on the InSb cores. Importantly, the potential to use these CdTe shells in hybrid superconducting-semiconducting nanowire devices is proposed based on the obtained high-quality InSb-CdTe interfaces and the small conduction band offset of this interface. In addition the comparable field-effect mobilities for both uncapped and CdTe-capped nanowires indicate the suitability of the CdTe shells to modulate the superconductor-semiconductor coupling without adding disorder to the device.
We note that we would have expected with the type-I band alignment of the InSb-CdTe interface, the nearly perfect epitaxy, and the high-quality interfaces an improvement in electron mobility in the CdTe-capped nanowires compared to the uncapped InSb wires. The similar mobility values in both uncapped and CdTe-capped nanowires likely suggests that thicker CdTe shells ($> 12$~nm) are required to confine the electron wavefunctions to the InSb core and attain higher mobility. While thicker shells are likely required for surface passivation, they are incompatible with tunnel-barrier devices since the tunneling probability decreases exponentially as a function of barrier thickness. Both functionalities\textemdash surface passivation and tunnel barriers\textemdash can be combined in a single nanowire device by growing asymmetrically thick shells, where the part of CdTe shell that will be in contact with the metal or superconductor is thin and the remaining nanowire facets are covered by a thick CdTe shell. Such hybrid nanowire devices that would combine reduced disorder\textemdash improved mobility\textemdash with tunable superconductor-semiconductor coupling could possibly open up an avenue for a new generation of topological nanowire devices.

\section*{Acknowledgements}
This work received support from the Dutch Organization for Scientific Research (NWO), the Foundation for Fundamental Research on Matter (FOM), the European Research Council (ERC HELENA 617256), and Microsoft Corporation Station-Q. Solliance and the Dutch province of Noord-Brabant are acknowledged for funding the TEM facility. S.M.F. is supported by the U.S. Department of Energy, Basic Energy Sciences grant DE-SC0022073 for transport measurements.
S.B. and T.R. acknowledge funding from the Volkswagen Stiftung (Momentum) through the project "dandelion'' and from the Deutsche Forschungsgemeinschaft (DFG, German Research Foundation) through the project BO 4280/8-1.
The Atom probe tomography work was supported by NSERC Canada (Discovery, SPG, and CRD Grants), Canada Research Chairs, Canada Foundation for Innovation, Mitacs, PRIMA Québec, and Defence Canada (Innovation for Defence Excellence and Security, IDEaS) and was performed at the Northwestern University Center for Atom-Probe Tomography (NUCAPT). NUCAPT is supported by the MRSEC program (NSF DMR-1720139), the SHyNE Resource (NSF ECCS-2025633), and instrumentation grants from the NSF-MRI (DMR-0420532) and ONR-DURIP (N00014-0400798, N00014-0610539, N00014-0910781, N00014-1712870) programs.


\section*{Data availability}
 The data that support the figures, the transport measurement plots and other findings are available at 10.5281/zenodo.5592058.



\section*{References}
\bibliography{ref_lib_e}

\end{document}


\vspace*{0ex}
	\title{Supplementary Information\\}
	\maketitle
	\thispagestyle{fancy}
	\fancyhf{}
	\rhead{}

\section{Density functional theory calculations} \label{SI_dft}
 All density functional theory (DFT) calculations are performed using the Vienna \textit{ab initio} simulation package (VASP)~\cite{kresse1996efficient}. These calculations employ the VASP implementation of the generalized Kohn-Sham scheme with the projector augmented-wave (PAW) method~\cite{blochl1994projector}. A plane-wave cutoff of 274.3~eV is used and spin orbit coupling is included in all calculations. While the results presented in the main text are obtained using the Heyd-Scuseria Ernzerhof (HSE06) hybrid exchange-correlation functional~\cite{heyd2003hybrid,err}, in this section results from additional exchange-correlation (xc) functionals are considered: the Perdew-Burke-Ernzerhof (PBE)~\cite{perdew1996generalized}, the modified Becke-Johnson (mBJ)~\cite{tran2009accurate}, and the local mBJ (lmBJ)~\cite{rauch2020local} (see Table~\ref{table_dft}).
 
 For the bulk calculations, the lattice constants of InSb and CdTe are set to the experimental values of 6.479~\AA~and 6.482~\AA, respectively~\cite{farrow1981molecular}.
  These calculations provide the bulk bandgaps ($E_\text{g}$) of both materials. For the supercell calculation, each material is represented by 10 monolayers, i.e., 40 atoms in total, connected at a non-polar (110) plane and the Brillouin zone is sampled using a $\Gamma$-centered 6~$\times$~6~$\times$~2 \textbf{k}-point grid. The tabulated valence and conduction band offsets \textemdash $\Delta E_\text{v}$ and $\Delta E_\text{c}$, respectively\textemdash are extracted using both the individual bulk calculations and the supercell calculation~\cite{borlido2018local}. In particular, from the supercell calculation, the potential offset, $\Delta V$, between InSb and CdTe is obtained using a macroscopic average of the electrostatic potential. This value is then used to align the energy levels obtained from the two separate InSb and CdTe bulk calculations. 
  
 \begin{table}[b!]
{\small
\setlength{\tabcolsep}{18pt}
\begin{tabular}{ l   r   r  r   r}
\textbf{XC functional}  &  \textbf{$\bm{E}_\text{g}$ (InSb)} &  \textbf{$\bm{E}_\text{g}$ (CdTe)} &  \textbf{$ \bm{\Delta} \bm{E}_\text{v}$ }  & \textbf{$ \bm{\Delta} \bm{E}_\text{c}$ }\\ 
\vspace{-0.05cm} & \vspace{-0.05cm} & \vspace{-0.05cm} & \vspace{-0.05cm} & \vspace{-0.05cm}\\
\hline
\vspace{-0.05cm} & \vspace{-0.05cm} & \vspace{-0.05cm} & \vspace{-0.05cm} & \vspace{-0.05cm}\\
PBE & 0.0 & 0.49 & -0.63 & -0.14 \\

mBJ & 0.24 & 1.57 & -0.65 & 0.68 \\

lmBJ & 0.21 & 1.54 & -0.68 & 0.65 \\

HSE06 & 0.28 & 1.27 & -0.84 & 0.15 \\

experimental & 0.24 & 1.45 & -0.87 & 0.34$^{*}$
    
      \end{tabular}
      }
\caption{\textbf{Bulk band structures and band offsets.} Values are given in electron Volts (eV). Experimental bandgaps are obtained from~\cite{vurgaftman2001band,chu1993recent,greene1990fundamental}.
The valence band offset is estimated from x-ray photoelectron spectroscopy in~\cite{mackey1986chemical}. The asterisk signifies that the experimental $\Delta E_\text{c}$ is calculated from the experimental bandgaps and $\Delta E_\text{v}$.}
\label{table_dft}
\end{table}

As shown in Table~\ref{table_dft}, the PBE functional predicts InSb to be a metal, i.e., with a zero bandgap, and thus this standard exchange-correlation functional cannot be used to describe the system. The mBJ, the lmBJ and, the HSE functionals give results which agree well with the experimental values. Specifically, HSE gives an accurate valence band offset compared to experiments, but it underestimates the bandgap of CdTe and accordingly $\Delta E_\text{c}$ by roughly 0.2~eV. Generally, the mBJ and the lmBJ exchange-correlation functionals are known to accurately describe bandgaps, while being numerically more feasible than hybrid functionals, which is also true for InSb and CdTe. However, both underestimate $\Delta E_\text{v}$ and correspondingly overestimate $\Delta E_\text{c}$ by approximately 0.3~eV, consistent with calculations in~\cite{hinuma2014band}.

\section{Atomic hydrogen cleaning} \label{ah cleaning}
Prior to the deposition of the CdTe shells, the native oxides surrounding the nanowires need to be removed to ensure epitaxy.
Before introducing the nanowire chips into the MBE system, they are glued to a molybdenum holder next to a temperature-reference chip of gallium arsenide (GaAs). The holder is then degassed for two hours at 300$^{\circ}$~C to ensure the desorption of any water molecules or unwanted adsorbates. Once degassed, the holder is inserted in the cleaning chamber. Table~\ref{table_aH} summarizes the relevant cleaning parameters. Once the oxide is removed, the nanowire chips are kept in the chamber until they have cooled down to 80$^{\circ}$~C and the chamber pressure has reached 3~$\times$~10$^{-9}$~Torr. The nanowire chips are then transferred through an ultra-high vacuum transfer tube to the growth chamber.

\begin{table}[b!]
{\small
\begin{tabular}{p{6cm} l  l}
\textbf{Parameter}  &  \textbf{Value}  \\ 
\vspace{-0.25cm} & \vspace{-0.25cm} \\
\hline
\vspace{-0.25cm} & \vspace{-0.25cm} \\

    Substrate temperature &  250$^{\circ}$~C  \\
    Hydrogen flow   & 20 sccm \\
    Filament temperature & 1200$^{\circ}$~C  \\
    Chamber pressure & 2.5~$\times$~10$^{-5}$~Torr  \\
    Cleaning duration & 45 minutes 
   \end{tabular} }
\caption{\textbf{Atomic hydrogen cleaning parameters.} The substrate temperature is measured on the surface of the temperature-reference chip. During the entire cleaning procedure the holder is rotated. When idle, the chamber pressure is roughly 6~$\times$~10$^{-10}$~Torr.}
\label{table_aH}
\end{table}

\begin{figure}[t!]
    \centering
    \includegraphics[scale=0.9]{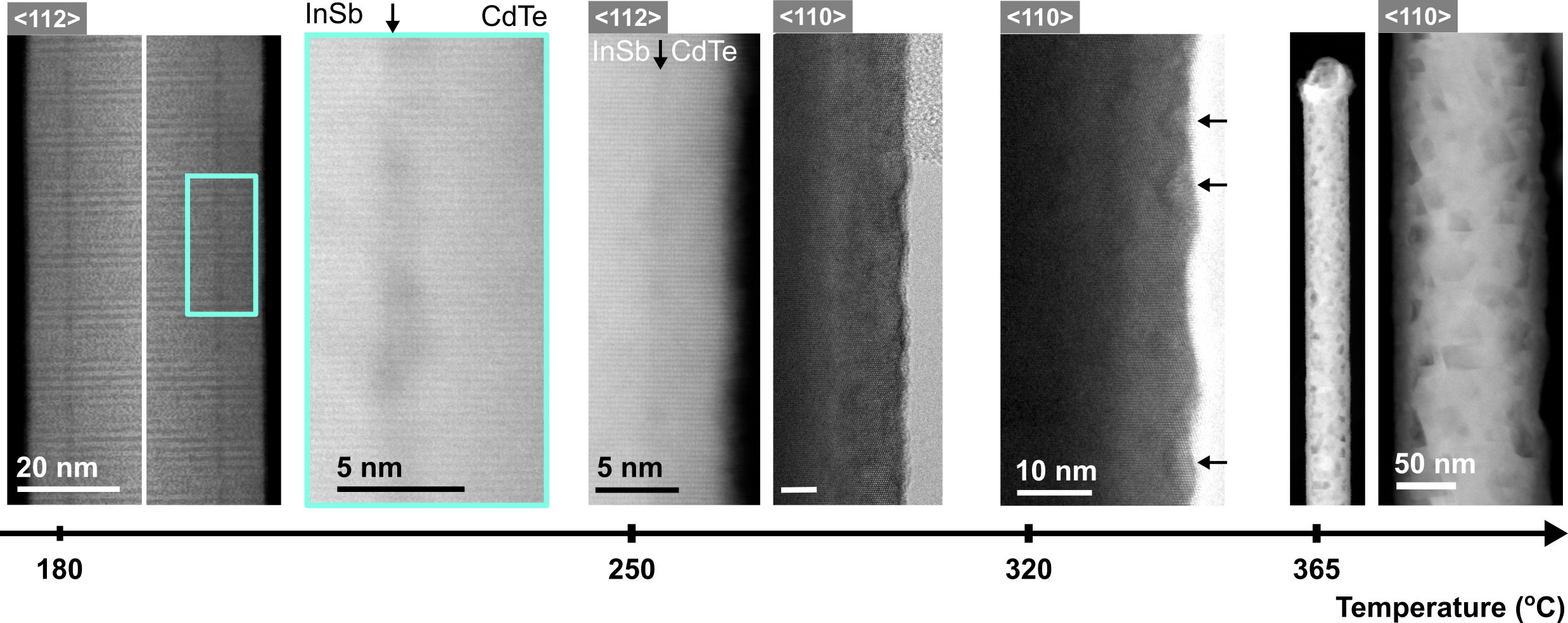}
    \caption{\textbf{Effect of substrate temperature during atomic hydrogen cleaning of core-shell nanowires.} Hydrogen cleaning of the InSb core at 180$^{\circ}$~C does not fully remove the oxide, as evident by the dark layer at the InSb-CdTe interface. A high magnification of the interface shows that this oxide layer is not of uniform thickness. At 250$^{\circ}$~C, this oxide layer is barely discernible and along the $<110>$ zone axis, the shell is defect-free signifying that the oxides are mostly removed. While 320$^{\circ}$~C instigates the onset of roughness in the InSb, at 365$^{\circ}$~C structural damage of the nanowire is detectable. In high-angle annular dark field imaging at both the $<112>$ and $<110>$ zone axes, damage manifests as pits on the nanowire surface. }
    \label{ah_fig}
\end{figure}

A series of cleaning times in combination with substrate temperatures have been investigated, as outlined in Figure~\ref{ah_fig}. While too high temperatures ($\, > 300^{\circ}$~C) induce roughness and damage the InSb nanowires surfaces, low temperatures are not sufficient to remove the native oxides. The remaining oxides manifest as a dark contrast at the InSb-CdTe interface in high-angle annular dark field (HAADF) imaging, reminiscent of a low electron-density material\textemdash~compared to InSb and CdTe. Accordingly, higher temperatures and longer cleaning times have been used to completely get rid of the oxide, since the cleanliness of the InSb surface determines to a great extent the quality of the grown CdTe shells~\cite{wood1984microstructural}.
Although the defect density in the CdTe shells is greatly minimized with optimized atomic hydrogen cleaning temperature and duration, implying that the native oxides are successfully removed, a dark interface contrast, though subtle, persisted regardless of cleaning conditions (Figure~\ref{ah_fig}: 250$^{\circ}$). The persistence of this interface layer leads us to conclude it is possibly arsenic (As) originating from the GaAs temperature-reference chip. In particular, during atomic hydrogen exposure, the GaAs chip is also being cleaned and is possibly releasing As at these temperatures which is being redeposited on the InSb nanowires. To substantiate the presence of As at the InSb-CdTe, atom probe tomography studies are used to analyze the nanowires since this interface layer is not detectable with elemental dispersive x-ray (EDX) spectroscopy.

\section{Atom probe tomography analysis} \label{apt}

\begin{figure}[t!]
    \centering
    \includegraphics[scale=1]{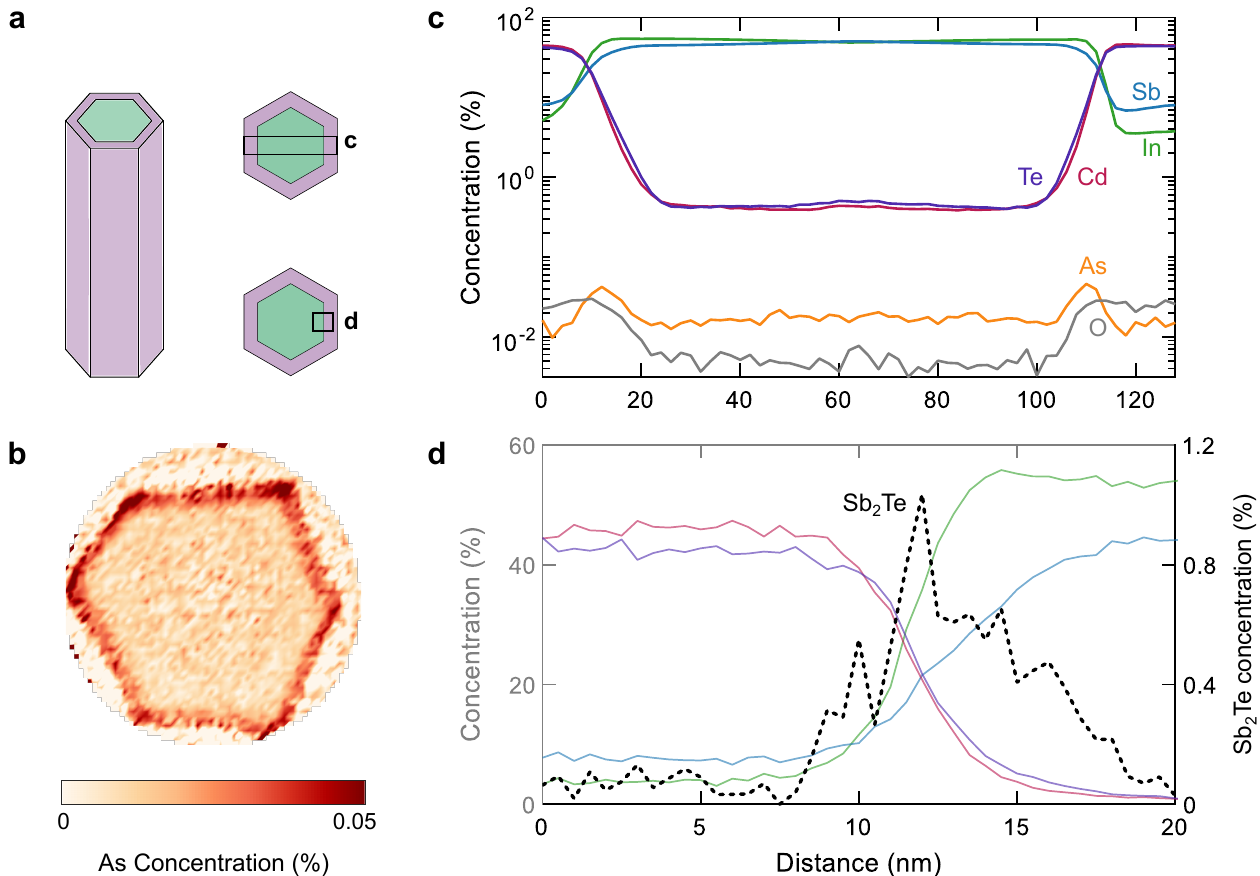}
    \caption{\textbf{Atom probe tomography analysis.}\textbf{ a.} A schematic of an InSb core (green) and a CdTe shell (purple) and two top-view images showing the analyzed profiles in \textbf{c.} and \textbf{d.}, a rectangular profile along the entire diameter and an interface profile, respectively.\textbf{ b.} Two-dimensional mapping of the arsenic concentration, reflecting a slight enrichment at the core-shell interface.\textbf{ c.} This arsenic concentration is also visible along the rectangular profile with two peaks at the interface of about 0.04\%. Roughly 0.03\% of oxygen is also present at the interface. The overlapping regions between the InSb and the CdTe are not due to interdiffusion but just peak overlaps. Within the detection resolution of roughly 0.1-1\% limited by these peak overlaps, there is no measurable interdiffusion.\textbf{ d.} An interface profile spanning a 20-nm region shows that Sb and Te are clustering up at the interface, forming a 3-4~nm thick layer. This layer contains Sb$_2$Te ions created during atom probe tomography indicating a mixed interface region/layer that incorporates both Sb and Te atoms rather than an abrupt interface between InSb and CdTe. }
    \label{apt_fig}
\end{figure}

For the atom probe tomography studies, the InSb nanowires were cleaned using atomic hydrogen at a substrate temperature of 250$^{\circ}$~C for 20 minutes followed by the growth of a 50-nm CdTe shell. These shells were grown with a Cd/Te ratio of 3 and a Cd pre-exposure time of one minute. These cleaning and growth conditions yield results akin to the one shown in Figure~\ref{ah_fig}: 250$^{\circ}$ with a very subtle dark contrast in high-angle annular dark field (HAADF) imaging.
Accordingly, we use the results from the atom probe tomography to fine-tune the growth conditions and to understand the origin of this darker interface layer, such that we can eliminate it. 
The results from the atom probe tomography analysis are presented in Figure~\ref{apt_fig} and confirm the presence of As at the InSb-CdTe interface. Moreover, very low levels of oxygen (approximately 0.03\%) are measured. Importantly, the dark interface layer is mostly attributed in this case to a tellurium-rich, possibly a Sb$_2$Te, layer. As shown in Figure~\ref{apt_fig}d, this Sb$_2$Te concentration extends into the InSb, confirming the presence of a Te-rich interface region.

To address the arsenic issue, the GaAs temperature-reference chip is covered with a 50~nm silicon nitride mask. The mask hinders the release of arsenic from the GaAs surface and thereby minimizes arsenic re-deposition on the nanowire surfaces during atomic hydrogen cleaning. With regards to the oxygen levels, the cleaning time is extended from 20 minutes to 45 minutes to ensure the complete removal of the native oxides. As for the Te-rich interface layer, its formation is suppressed by flushing the growth chamber with Cd before introducing Te, as detailed in the main text. These optimizations in the cleaning and growth parameters yield clean, smooth and abrupt InSb-CdTe interfaces evidenced by epitaxial shells absent of dark-contrast interfacial layers as discussed in the main text.

\section{CdTe growth} \label{SI_cdte}

\begin{figure}[t!]
    \centering
    \includegraphics[scale=1]{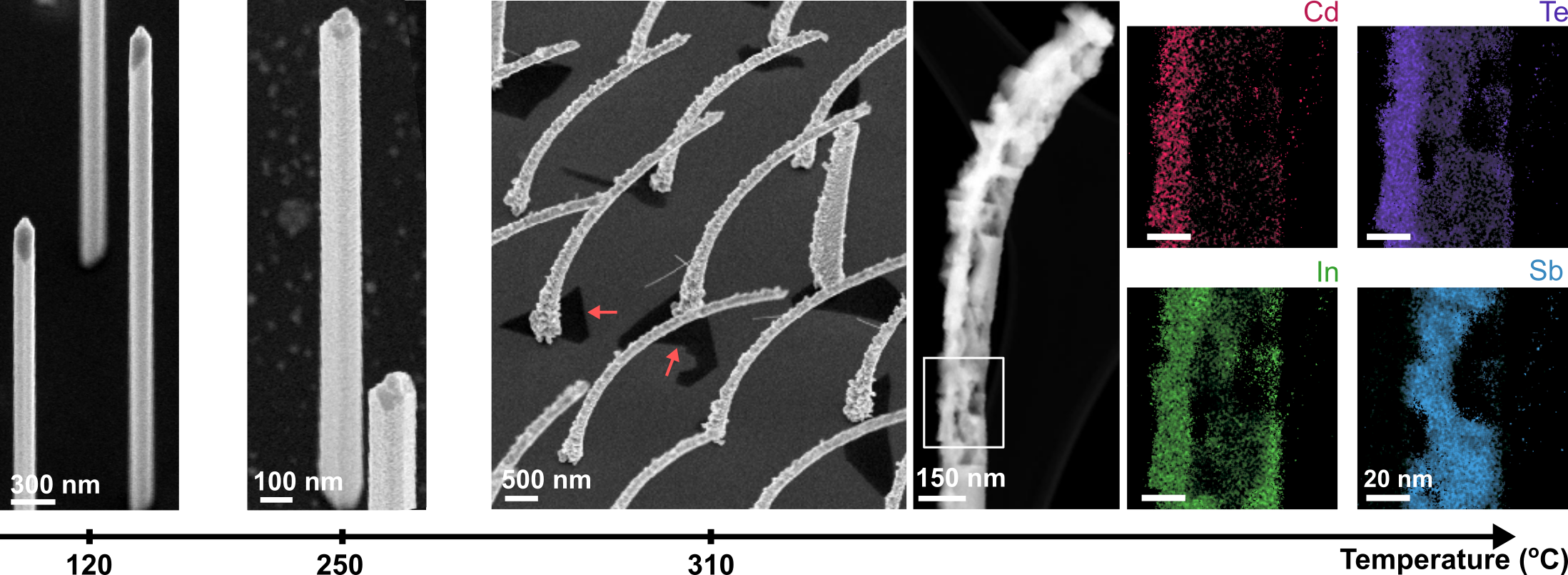}
    \caption{\textbf{CdTe shell growth at different temperatures.} Shell growth at 120$^{\circ}$~C yields smooth shells and a complete layer on the mask. At 250$^{\circ}$~C, roughness is already detectable within the resolution of the scanning electron microscope. Parasitic growth on the substrate reflects an increased diffusion length, resulting in islands rather than a complete layer. Even more roughness is induced at 310$^{\circ}$~C in addition to thermal etch pits in the substrate (tiny red arrows) and the nanowires are bent. A close examination of a single nanowire shows that the bending is instigated by pits in the nanowires. These voids are additionally visible in elemental dispersive x-ray mapping. The clustering of In towards the shell (overlapping with Te) signifies that possibly In$_2$Te$_3$ reactions took place at these temperatures.}
    \label{cdte-temp_fig}
\end{figure}

\normalsize{\textbf{Substrate temperature}}\\
The growth of CdTe shells takes place at relatively low temperatures, e.g., 120$^{\circ}$~C, since higher temperatures ($\approx 250^{\circ}$) result in rough and defected shells (Figure~\ref{cdte-temp_fig}) and are known to promote interface reactions between the InSb and the CdTe.
Even higher temperatures ($\,> 300^{\circ}$~C) compromise the structural integrity of the InSb nanowires. As shown in Figure~\ref{cdte-temp_fig}, high temperatures instigate the liberation of Sb from the nanowires and the InSb substrate. This release of Sb manifests as pits in both the nanowires and the substrate, causing the nanowires to bend. These higher temperatures increase the adatom surface-diffusion length, evident by the absence of deposition on the masked substrate surface. Conversely, at the optimal growth temperature of 120$^{\circ}$~C, the substrate surface is covered with a CdTe layer.

\section{Strain mapping} \label{strain}
\begin{figure}[t!]
    \centering
    \includegraphics[scale=1]{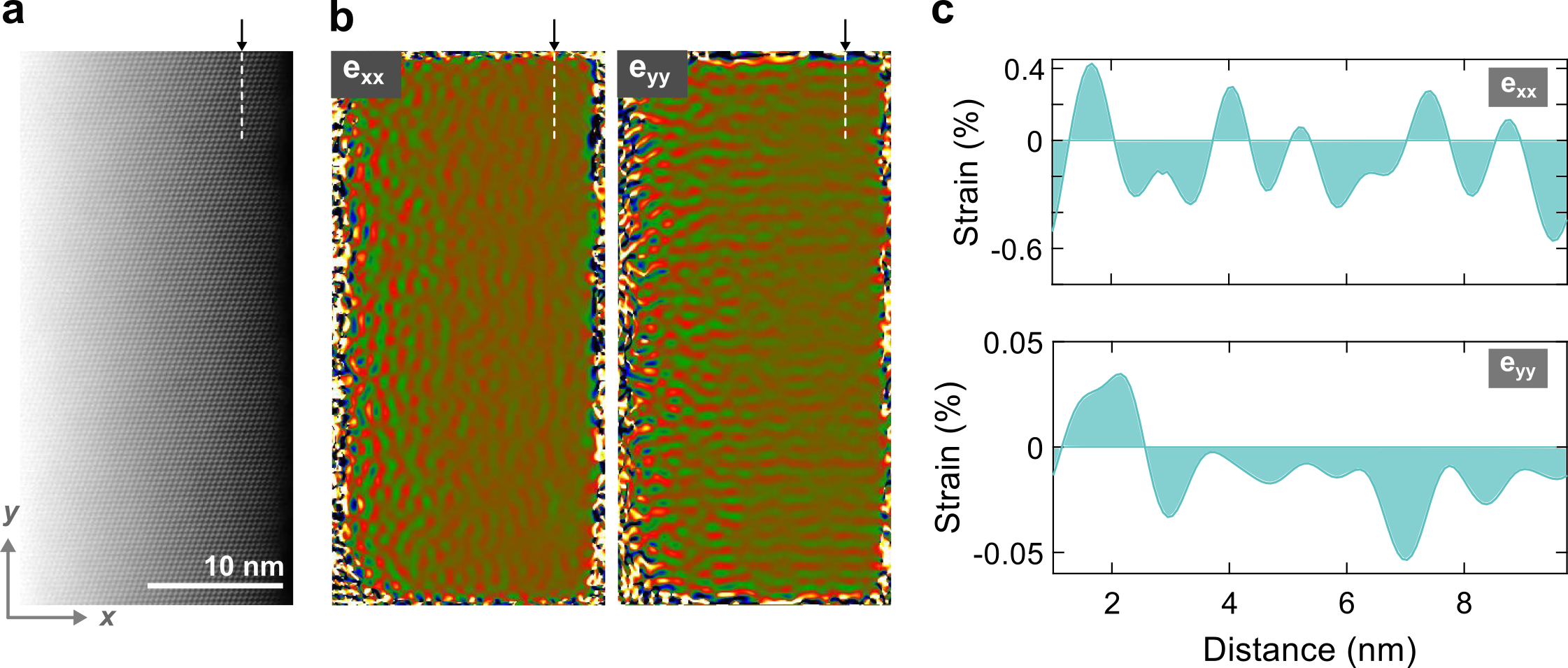}
    \caption{\textbf{Strain mapping.}\textbf{ a.} HAADF-STEM image of an InSb-CdTe nanowire taken along the $<110>$ zone axis with the interface indicated by an arrow.\textbf{ b.} The strain mapping of lattice spacing differences along the x and y directions for the HAADF-STEM image shown in\textbf{ a.} indicates a strain-free interface.\textbf{ c.} Line profiles of the strain maps in\textbf{ b.} show signal fluctuations below 0.5\% from the average, thus confirming the absence of strain. The indiscernible InSb-CdTe interface in the strain maps further substantiates a relaxed and epitaxial interface.}
    \label{strain_}
\end{figure}

 The InSb-CdTe core-shell nanowires are characterized to assess whether the interface is under strain. For this, an atomic resolution HAADF-STEM image is used where two $<-111>$ reflections in the fast Fourier transform (FFT) diffraction patterns are selected. The strain maps along the x and y directions, the e$_{xx}$ and e$_{yy}$ images respectively, do not show any clear edges at the interface and the signal fluctuation is less than 0.5\% from the average (Figure~\ref{strain_}). There is thus no clear indication of strain, within the detection limit of this technique. The absence of strain is consistent with the almost perfect lattice match between InSb and CdTe.
 
\section{Nano-scale roughness along the $<110>$ zone axis with increasing shell thickness}
\begin{figure}[t!]
    \centering
    \includegraphics[scale=1]{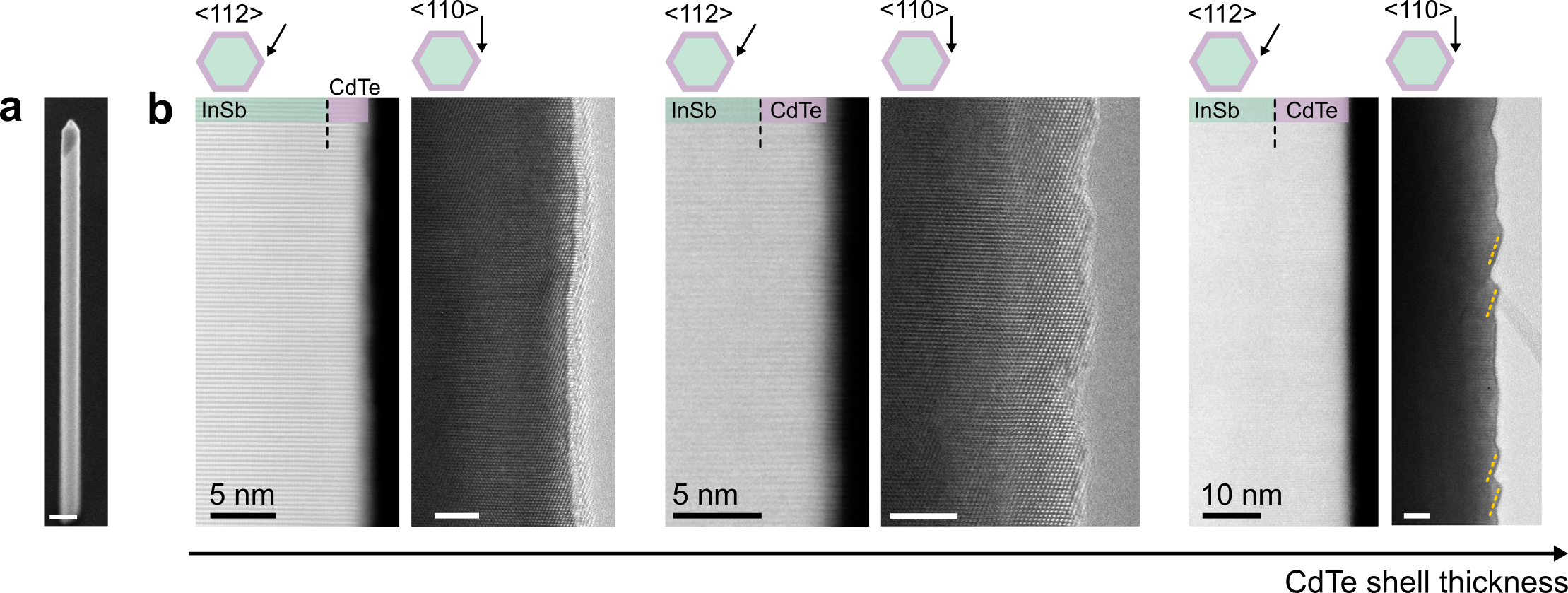}
  
    \caption{\textbf{Propensity to form nanofacets with increased CdTe thickness.}\textbf{ a.} An SEM image of single InSb nanowire covered with a CdTe shell. Scale bar is 200 nm.\textbf{ b.} HAADF scans and bright-field TEM images taken along the specified zone axes of nanowires with differently thick CdTe shells of 2.7 nm, 4 nm and 12.5 nm, respectively. Along the $<112>$ zone axis, the CdTe shells appear atomically flat. Along the $<110>$ direction, however, slight roughness is discernible and develops into well-defined non-vertical edges for the thickest shells. The roughly 13~nm shell is most likely terminated by \{111\} planes, as indicated by the yellow dashed lines.}
    \label{111planes_fig}
\end{figure}

Tuning the CdTe shell thickness is simply achieved by varying the growth time. For an increasing shell thicknesses (greater than 5~nm) very slight roughness is observed in the shell along the $<110>$ zone axis with transmission electron microscopy (Figure~\ref{111planes_fig}). Imaging the same shell along the $<112>$ zone axis does not reveal this roughness, since the nanowire is imaged parallel to a roughly 100~nm long nanowire facet, where a summation of nanoscale roughness is projected in the image plane. In contrast, in the $<110>$ zone axis the nanowire is viewed at the corner between two facets, thereby exposing any atomic scale roughness. This roughness shows up in projection at edges orthogonal to a $<111>$ direction. Although the exact topography and features of this roughness cannot be extracted, it could be due to an increased tendency of CdTe to form \{111\} facets with increased layer thickness, as already disclosed~\cite{neretina2006role,ribeiro2007low}. 

\newpage

\section{Cool-down under Te flux}
\begin{figure}[t!]
    \centering
    \includegraphics[scale=1]{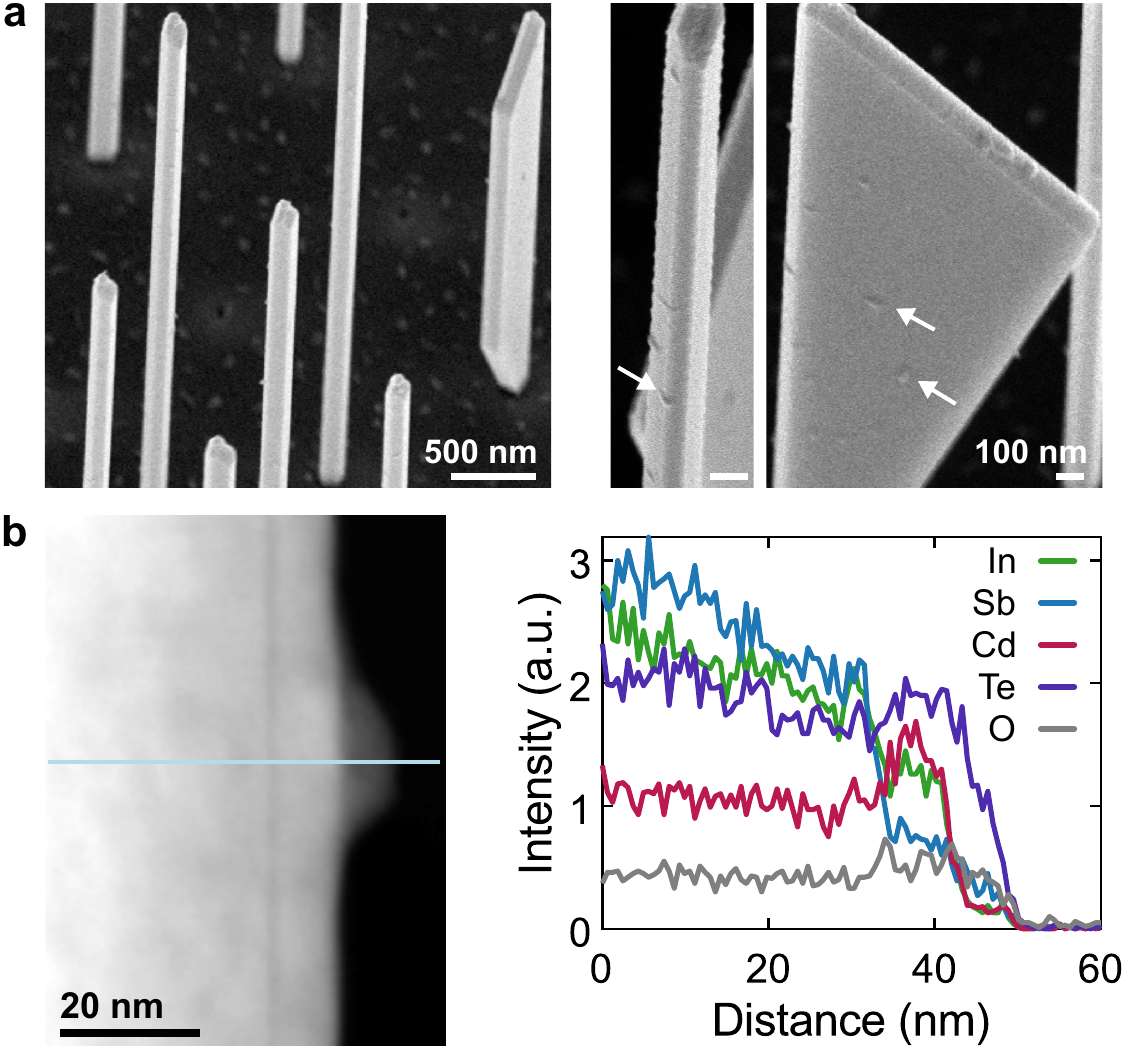}
    \caption{\textbf{Cool-down under Te flux.}\textbf{ a.} SEM images taken at 30$^{\circ}$-tilt showing tiny globules on the substrate surface and on top of the CdTe shells. Arrows point to representative globules on a nanowire and a nanoflake (scale bar: 100~nm).\textbf{ b.} HAADF-STEM image displaying an area with a globule. An EDX line scan is acquired along the blue line. The composition profile indicates the globule has a higher Te content compared to the CdTe shell.}
    \label{cdte-cooldown_fig}
\end{figure}

CdTe shell growth is terminated by closing both the Cd and Te shutters. However, for experiments in which a Te flux is supplied during cool-down of the substrate for 15 minutes, Te-rich CdTe globules are deposited on the CdTe shell (Figure~\ref{cdte-cooldown_fig}). Conversely, for shells grown at higher temperatures (200$^{\circ}$~C) these globules are missing, likely related to a longer diffusion length.

\newpage

\newpage
\section{Device fabrication details} \label{device_fabrication}
 
The fabrication process for mobility devices is described:
\begin{itemize}
    \item p$^{++}$-doped Si substrates are covered by 285~nm SiO$_x$ via dry oxidation and 15-nm hafnium oxide (HfO$_x$) via atomic layer deposition, onto which titanium/gold alignment markers are defined.
    
    \item The nanowires are deterministically transferred by a micro-manipulator from the growth chip onto the global back-gate chips. SEM images allow for designing the devices with respect to the alignment markers.
    
    \item The e-beam resist, PMMA 950 A4 is spun at 6000 rpm for 1 minute onto the chips, after which it is baked at 175$^{\circ}$~C for 15 minutes.
    
    \item The pattern of the contacts is defined by electron-beam lithography on the resist. Source and drain contacts are separated by InSb channels of lengths 1, 2, 3, and 5~$\mu$m.
    
    \item The chips are developed in a solution of MIBK:IPA (1:3) for 1 minute and in IPA for 1 minute.
    
    \item Prior to the metal deposition, the chips are cleaned for 30 seconds with a 200-Watt oxygen plasma and a flow of oxygen equal to 78 sccm. 
    
    \item The substrate is loaded into an e-gun evaporator. The CdTe shell is removed via Ar milling set to 250~V and 15~mA for 3 minutes (for the 4-nm CdTe shell). This duration is divided into 30-second intervals to avoid over-heating of the nanowires. 
    
    \item In the same chamber, a 10-nm layer of titanium followed by a 150-nm layer of gold are evaporated to form the contacts.
        
    \item The resist is lift-off in acetone at room temperature overnight.
\end{itemize}
 
\newpage
 \section{Mobility measurements} \label{transport_results}

For the study of the field-effect mobility $\mu$, nanowire field-effect transistor (FET) devices were fabricated as described in Section~\ref{device_fabrication}. Based on the channel lengths $L$ used, the diffusive long-channel regime is assumed. Hence, the current $I$ as a function of back-gate voltage $V_{\text{BG}}$ can be modeled by 
\begin{equation}
I (V_\text{BG}) = \frac{V_\text{dc}}{(L^2/\mu C) (V_\text{BG} - V_\text{th})^{-1} + R_\text{c}},
\end{equation}
where~$V_\text{dc}$ is the bias voltage. The saturation current is limited by the series resistance $R_{\text{c}}$, which includes the contact, filter, and line resistances. The current pinch-off is reached at the threshold voltage $V_{\text{th}}$. The value of the capacitance~$C$ is evaluated via a 3D Laplace solver for a typical nanowire device geometry, where the core-shell nanowire diameter is assumed to be 120-140~nm and the 15-nm hafnium oxide layer is accounted for. In this finite-elements model the InSb nanowire is treated as a metal~\cite{gul2015towards}. The capacitance values for bare InSb nanowires and different CdTe shell thicknesses are given in Table~\ref{table_cap}.

  \begin{table}[b!]
 {\small
 \setlength{\tabcolsep}{19pt}
 \begin{tabular}{l l l l l l}
  \null &  &  \multicolumn{4}{c}{\textbf{Channel length}} \\ 
 \vspace{-0.05cm} & \vspace{-0.05cm}  &\vspace{-0.05cm} &\vspace{-0.05cm} & \vspace{-0.05cm} &  \vspace{-0.05cm}\\

 \textbf{CdTe thickness} &  &  \textbf{1~$\bm{\mu}$m} &  \textbf{2~$\bm{\mu}$m}   & \textbf{3~$\bm{\mu}$m} & \textbf{5~$\bm{\mu}$m}\\ 

 \vspace{-0.04cm} & \vspace{-0.04cm}  & \vspace{-0.04cm} &\vspace{-0.04cm} & \vspace{-0.04cm} &  \vspace{-0.04cm}\\
 \hline
 \vspace{-0.04cm}  & \vspace{-0.04cm} & \vspace{-0.04cm} &\vspace{-0.04cm} & \vspace{-0.04cm} &  \vspace{-0.04cm}\\
 0~nm & & $36.8$ & $85.8$ & $134.8$ & 232.8\\
 \vspace{-0.025cm} &   \vspace{-0.025cm} & \vspace{-0.025cm}& \vspace{-0.025cm} & \vspace{-0.025cm}  &  \vspace{-0.025cm} \\
 2~nm & & $37.9$ & $87.8$ & $137.9$ & 238.1 \\
 4~nm & & $37.2$ & $86.6$ & $136.0$ & 235.1 \\
 7~nm & & $37.1$ & 86.5 & $135.9$ &  235.0\\
 12~nm & & $36.6$ & $85.6$ & $134.8$ & 233.1 
  
      \end{tabular}
      }
 \caption{\textbf{Capacitance values used for mobility fitting.} The device capacitance is evaluated for different channel lengths and CdTe shell thicknesses. Capacitance values are given in aF.}
 \label{table_cap}
 \end{table}

\begin{figure}[t!]
    \centering
\includegraphics[scale=1]{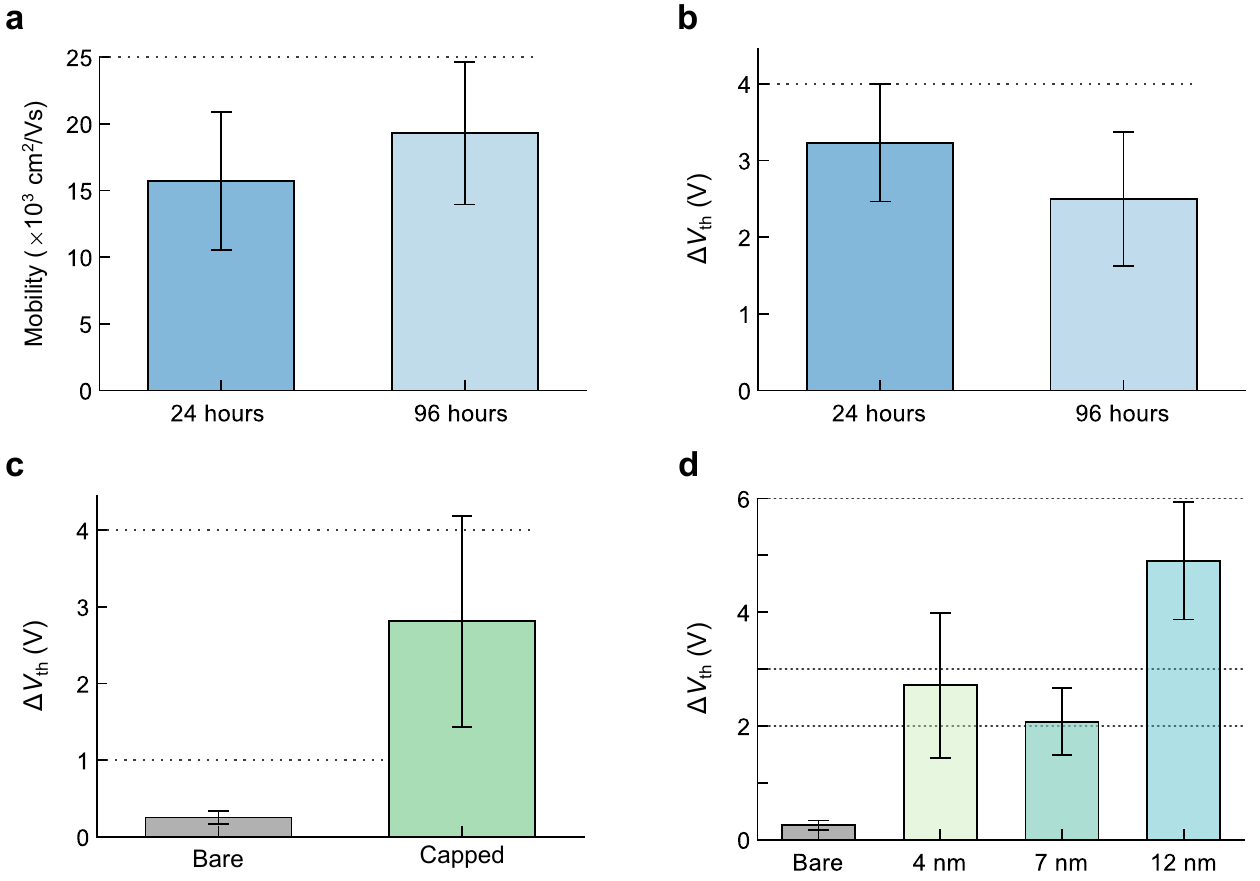}
    \caption{\textbf{Effect of sample space evacuation on mobility and hysteresis.}\textbf{ a.} and\textbf{ b.} FET nanowire devices with a 4-nm CdTe shell are evacuated for 24 and 96 hours.\textbf{ a.} A longer evacuation leads to a slight mobility enhancement going from 1.57$\times 10^{4}$~cm$^2$/Vs to 1.93$\times 10^{4}$~cm$^2$/Vs.\textbf{ b.} A decrease in hysteresis from $\Delta V_{\text{th}}$ = 3.21~V to 2.49~V, is noted for a 96-hour evacuation, compared to 24 hours. Hysteresis is quantified by the threshold-voltage difference $\Delta V_{\text{th}}$ between the forward and backward gate-voltage sweeps.\textbf{ c.}~CdTe-capped wires show a much larger hysteresis compared to uncapped, bare InSb nanowires.\textbf{ d.} Hysteresis varies with CdTe shell thickness, with the largest hysteresis present for the 12-nm CdTe shell nanowires.}
    \label{hyst}
\end{figure}

All FET nanowire devices for mobility extraction are measured in a dip-stick in helium at T = 4.2 K. Prior to cool down, the sample space is evacuated at room temperature for 24, 48 and 96 hours to efficiently desorb adsorbates from the nanowire surface. In Figure~\ref{hyst}a-b, devices that have been pumped for 24 and 96 hours are presented to evaluate the effect of sample space evacuation on device performance. Particularly, a slight increase in mobility, going from 1.57$\times 10^{4}$~cm$^2$/Vs to 1.93$\times 10^{4}$~cm$^2$/Vs, is noted for the devices evacuated for 96 hours compared to 24 hours (Figure~\ref{hyst}a). The data depicted in Figure~\ref{hyst}a is for devices with a 4-nm CdTe shell. Longer sample space evacuation also reduces the measured hysteresis between the forward and backward back-gate voltage sweeps for the CdTe-capped wires, as illustrated in Figure~\ref{hyst}b. The hysteresis is quantified by determining the difference in threshold voltages ($\Delta V_{\text{th}}$) between both sweep directions. The devices studied in Figure~\ref{hyst}b, indicate that on average hysteresis decreased from 3.21~V to 2.49~V as function of longer sample evacuation.

The presence of such a large hysteresis is associated with the CdTe shells, since hysteresis is nearly absent in uncapped, bare InSb nanowires (Figure~\ref{hyst}c). Although the mechanism by which the CdTe shell enhances the hysteresis remains unknown, it is likely attributed to point defects in the CdTe shells and adsorbates related to these point defects. The slight improvement in the hysteresis for longer sample space evacuation suggests that something inherent to the shell is the source of this hysteresis. As depicted in Figure~\ref{hyst}d, the largest hysteresis exists for the thickest studied CdTe shells (12~nm). This observation likely suggests that a larger CdTe volume hosts more point defects in contrast to thinner shells. 

\clearpage

\Large{\textbf{References}}

\bibliography{ref_lib_e}